\documentclass[prl,10pt,aps,twocolumn,superscriptaddress,nofootinbib,showpacs,floatfix,amssymb]{revtex4-1}
\usepackage{slashed}
\usepackage{epsfig}

\newcommand{\beqn}{\begin{eqnarray}}
\newcommand{\eeqn}{\end{eqnarray}}
\newcommand{\eq}[1]{(\ref{#1})}

\begin{document}

\title{Vafa-Witten theorem, vector meson condensates and magnetic--field--induced electromagnetic superconductivity of vacuum}

\author{M. N. Chernodub}\email{On leave from ITEP, Moscow, Russia.}
\affiliation{CNRS, Laboratoire de Math\'ematiques et Physique Th\'eorique, Universit\'e Fran\c{c}ois-Rabelais Tours,\\ F\'ed\'eration Denis Poisson, Parc de Grandmont, 37200 Tours, France}
\affiliation{Department of Physics and Astronomy, University of Gent, Krijgslaan 281, S9, B-9000 Gent, Belgium}

\begin{abstract}
We show that the electromagnetic superconductivity of vacuum in strong magnetic field background is consistent with the Vafa-Witten theorem because the charged vector meson condensates lock relevant internal global symmetries of QCD with the electromagnetic gauge group.
\end{abstract}

\pacs{13.40.-f, 12.38.-t, 74.90.+n}


\date{September 14, 2012}

\maketitle

The Vafa-Witten (VW) theorem shows that vector-like global symmetries (like isospin or baryon number) cannot be spontaneously broken in vector-like theories (for example, in QCD) with zero theta angle~\cite{Vafa:1983tf}. According to Ref.~\cite{Hidaka:2012mz}, a generalization of this theorem implies that vector meson condensation cannot occur in QCD vacuum in the background of a strong magnetic field because these condensates would inevitably break certain internal global symmetries of QCD and, consequently, lead to appearance of massless Nambu-Goldstone boson associated with this symmetry breaking. 

The presence of the charged vector condensates is crucial for the magnetic-field-induced electromagnetic superconductivity of vacuum which appears in certain hadronic models coupled to electromagnetism~\cite{Chernodub:2010qx}. A sufficiently strong magnetic field, $eB \sim 1\, \mathrm{GeV}^{2}$, may lead to the condensation of the charged vector mesons
\beqn
\rho_{\pm} (x) = \langle \bar \psi(x) \gamma_{\pm} \tau_{\pm} \psi(x) \rangle\,, 
\label{eq:rho}
\eeqn
where $\psi = (u,d)^{T}$ is the quark spinor in Dirac, color and flavor space, $\gamma_{\pm} = (\gamma_{1} \pm i \gamma_{2})/2$ and $\tau_{\pm} = (\tau_{1} \pm i \tau_{2})/2$ are combinations of the Dirac (spinor) and Pauli (flavor) matrices, respectively. Below we demonstrate --  contrary to the statement of Ref.~\cite{Hidaka:2012mz} -- that the presence of the nonvanishing condensates~\eq{eq:rho} in QCD in the  magnetic field background is consistent with the VW theorem.

The Lagrangian of two-flavor QCD in the background of the electromagnetic field $A_{\mu}^{\mathrm{em}}$ reads as follows:
\beqn
{\cal L} = - \frac{1}{4} G^{a}_{\mu\nu} G^{a\mu\nu} - \bar \psi (i \gamma^{\mu} D_{\mu} - m) \psi\,,
\label{eq:L}
\eeqn
where $G^{a}_{\mu\nu}$ is the strength tensor of the gluon field $A^{a}_{\mu}$,
\beqn
D_{\mu} = \partial_{\mu} - i g T^{a} A^{a}_{\mu} - i q A_{\mu}^{\mathrm{em}}\,,
\eeqn
is the covariant derivative, $T^{a}$ are generators of the $SU(3)_{c}$ color group and
\beqn
q = \frac{e}{2} \bigl(\tau_{3} + \frac{1}{3}\bigr)\,,
\label{eq:q}
\eeqn
is the electric charge matrix acting in the flavor space. For simplicity, the masses of up and down quarks are taken to be the same, $m_{u} = m_{d} = m$.  

Due to the difference in electric charges of up and down quarks~\eq{eq:q}, $q_{u} = 2 e/3$ and $q_{d} = - e/3$, the group of the internal continuous global symmetries of Lagrangian~\eq{eq:L} is explicitly broken by the background electromagnetic field $A_{\mu}^{\mathrm{em}}$:
\beqn
SU_{V}(2) \times U_{B}(1) \to U(1)_{I_{3}} \times U(1)_{B}\,,
\label{eq:global:remaining}
\eeqn
where $U(1)_{I_{3}}$ is the diagonal subgroup of the isospin group $SU(2)_{V}$ and $U(1)_{B}$ is the baryon number symmetry. 

The internal local symmetries of Lagrangian~\eq{eq:L} include the electromagnetic $U(1)_{\mathrm{em}}$ gauge symmetry
\beqn
U(1)_{\mathrm{em}}: \quad 
\left\{\begin{array}{rcl}
A_{\mu}^{\mathrm{em}}(x) & \to & A_{\mu}^{\mathrm{em}}(x)  + \partial_{\mu} \omega_{\mathrm{em}}(x)\\
\psi(x) & \to & \exp\{i \omega_{\mathrm{em}} (x) q \} \psi(x)
\end{array}
\right.\,,
\label{eq:Uem}
\eeqn
and the color $SU(3)_{c}$ gauge symmetry. These local symmetries are not anomalously broken so that the fermion determinant is invariant under the local $SU(3)_{c} \times U(1)_{\mathrm{em}}$ gauge group. The background magnetic field itself does not break explicitly the electromagnetic gauge symmetry~\eq{eq:Uem} since the magnetic field is defined by a component of the gauge invariant Abelian field strength tensor $F^{\mathrm{em}}_{\mu\nu} = \partial_{\mu} A^{\mathrm{em}}_{\nu} - \partial_{\nu} A^{\mathrm{em}}_{\mu}$. Thus the Abelian symmetry~\eq{eq:Uem} is the symmetry of QCD in the background of magnetic field~\eq{eq:L} regardless if the background magnetic field is a quantized (dynamical) field or a classical (static) field.  

The condensates~\eq{eq:rho} are obviously invariant under the baryonic $U(1)_B$ transformations, $\psi \to e^{i \omega_B} \psi$, while the remaining global $U(1)_{I_{3}}$ group,
$\psi \to e^{i \omega_{{I_{3}}} \tau_{3}/2}\, \psi$, transforms the vector condensates~\eq{eq:rho} as follows:
\beqn
U(1)_{I_{3}}: \quad \rho_{\pm}(x) \to e^{ \mp i \omega_{I_{3}}}\rho_{\pm}(x) \,.
\label{eq:U1:I3}
\eeqn

It was noted in Ref.~\cite{Hidaka:2012mz} that a possible spontaneous breaking of the $U(1)_{I_{3}}$ global symmetry~\eq{eq:U1:I3} by the vector condensates~\eq{eq:rho} contradicts the VW theorem. The spontaneous breaking of the global symmetry should give rise to appearance of a massless Nambu-Goldstone boson. However, the corresponding massless boson does not emerge in the superconducting phase of QCD because the $U(1)_{I_{3}}$ global transformation~\eq{eq:U1:I3} is, in fact, a part of the larger, electromagnetic symmetry group~\eq{eq:Uem}:
\beqn
U(1)_{\mathrm{em}}:  \quad \rho_{\pm}(x) \to e^{ \mp i \omega_{\mathrm{em}}(x)}\rho_{\pm}(x) \,.
\label{eq:U1:rho}
\eeqn

Equation~\eq{eq:U1:rho} reflects the trivial fact that the vector quantities~\eq{eq:rho} are condensates of the electrically charged particles, so that they are sensitive to the electromagnetic $U(1)_{\mathrm{em}}$ transformation~\eq{eq:Uem} as well. Thus, the condensates~\eq{eq:rho} break the internal local symmetry~\eq{eq:Uem} and in this case the Nambu-Goldstone boson is known to be absent~\cite{ref:Weinberg:book} in agreement with the VW theorem~\cite{Vafa:1983tf}: the would-be Nambu-Goldstone boson is absorbed into the Abelian gauge field $A_{\mu}^{\mathrm{em}}$ thus making the photon massive (as it happens in the ordinary superconductivity~\cite{ref:Weinberg:book}).

Basically, the action~\eq{eq:U1:I3} of the subgroup $U(1)_{I_{3}}$ of the global $SU(2)_{V}$ group on the condensates~\eq{eq:rho} is identical to (or, ``locked with'') the action of a global subgroup of the electromagnetic $U(1)_{\mathrm{em}}$ gauge group~\eq{eq:U1:rho} on the same condensates. 

Another argument against the breaking of the global $U(1)_{I_{3}}$ internal symmetry in the superconducting phase of QCD is as follows. The positive, $\rho^{+}$, and negative, $\rho^{-}$, condensates~\eq{eq:rho} -- which are, strictly speaking, independent quantities -- appear simultaneously in the superconducting phase of the vacuum. These condensates  have the same absolute values and opposite phases~\cite{Chernodub:2010qx} stressing the fact that the superconducting vacuum is an electrically neutral state. Therefore, the vacuum state is annihilated by the conserved electric charge operator. Since the generators of the $U(1)_{\mathrm{em}}$ and $U(1)_{I_{3}}$ groups are the same modulo the global $U(1)_{B}$ baryon symmetry (which is never broken anyway), the $U(1)_{I_{3}}$ charge operator annihilates the vacuum state as well. The latter fact highlights the absence of the breaking of the global internal $U(1)_{I_{3}}$ symmetry in the superconducting QCD phase.

As a side remark, it is worth mentioning that analytical derivations of Vafa-Witten-type theorems may contain various technical loopholes which depend on particular physical circumstances. A similar Vafa-Witten theorem on parity violating condensates~\cite{Vafa:1984xg}, for example,  ``cannot be regarded as an established mathematical theorem''~\cite{Kanazawa:2011tt}. In other words, if under certain (external) conditions the physical arguments allow for the presence of certain ``VW-forbidden'' condensates then the analytical proof of the VW theorem turns out to be invalid due to technical reasons related to specific features of these physical conditions. The relevant examples related to flavor and/or parity violating condensates are discussed in the context of finite temperature QCD~\cite{Cohen:2001hf}, lattice regularized QCD with Wilson fermions~\cite{Sharpe:1998xm}, in QCD at finite isospin potential~\cite{Son:2000xc} etc. 

Finally, we would like to stress an important difference between spontaneous breaking of external and internal continuous symmetries\footnote{A discussion of the ``locking'' of the external and internal symmetries of QCD by the condensates~\eq{eq:rho} may be found in~\cite{Chernodub:2010qx}.}. The spectrum of the superconducting phase of QCD should contain the massless bosonic modes associated with breaking of certain {\emph {external}} (rather than {\emph{internal}}) global symmetries of Lagrangian~\eq{eq:L}. These modes are related not to the fact of the very presence of the superconducting condensates~\eq{eq:rho} and not to the internal global symmetries of QCD, but rather to the spatial (``external'') inhomogeneities of the condensates~\eq{eq:rho} in the superconducting ground state of the theory. The inhomogeneous condensates possess the vortex lattice structure which breaks spontaneously translational and rotational symmetries of the vacuum~\cite{Chernodub:2010qx} in a complete analogy with the mixed Abrikosov phase of the type--II superconductors~\cite{ref:Weinberg:book,ref:review:vortices}. The spontaneous breaking of these external symmetries leads to appearance of the so called ``supersoft Goldstone shear modes'' related to vibrations of the vortex lattice in the ground state of the theory~\cite{ref:review:vortices}. 

Summarizing, we have shown that the electromagnetic superconductivity of vacuum in strong magnetic field background is consistent with the Vafa-Witten theorem.

\begin{acknowledgments}
The author is grateful to Y.~Hidaka and A.~Yamamoto for useful correspondence. The work was supported by Grant No. ANR-10-JCJC-0408 HYPERMAG (France). 
\end{acknowledgments}


\begin{thebibliography}{99}

\bibitem{Vafa:1983tf} 
  C.~Vafa and E.~Witten,
  Nucl.\ Phys.\ B {\bf 234}, 173 (1984).

\bibitem{Hidaka:2012mz} 
  Y.~Hidaka and A.~Yamamoto,
  arXiv:1209.0007 [hep-ph].

\bibitem{Chernodub:2010qx} 
  M.~N.~Chernodub,
  Phys.\ Rev.\ D {\bf 82}, 085011 (2010)
  [arXiv:1008.1055 [hep-ph]];
  Phys.\ Rev.\ Lett.\  {\bf 106}, 142003 (2011)
  [arXiv:1101.0117 [hep-ph]];
 M.~N.~Chernodub, J.~Van Doorsselaere and H.~Verschelde,
  Phys.\ Rev.\ D {\bf 85}, 045002 (2012)
  [arXiv:1111.4401 [hep-ph]].

\bibitem{ref:Weinberg:book}
S. Weinberg, The Quantum Theory of Fields, Vol. II
(Cambridge University Press, Cambridge, UK, 1996).

\bibitem{ref:review:vortices}
B.~Rosenstein and D.~Li, 
Rev. Mod. Phys. {\bf 82}, 109 (2010).

\bibitem{Vafa:1984xg} 
  C.~Vafa and E.~Witten,
  Phys.\ Rev.\ Lett.\  {\bf 53}, 535 (1984).

\bibitem{Kanazawa:2011tt} 
  T.~Kanazawa, T.~Wettig and N.~Yamamoto,
  JHEP {\bf 1112}, 007 (2011)
  [arXiv:1110.5858 [hep-ph]].

\bibitem{Cohen:2001hf} 
  T.~D.~Cohen,
  Phys.\ Rev.\ D {\bf 64}, 047704 (2001)
  [hep-th/0101197].

\bibitem{Sharpe:1998xm} 
  S.~Aoki and A.~Gocksch,
  Phys.\ Rev.\ D {\bf 45}, 3845 (1992);
  S.~R.~Sharpe and R.~L.~Singleton, Jr,
  Phys.\ Rev.\ D {\bf 58}, 074501 (1998)
  [hep-lat/9804028].

\bibitem{Son:2000xc} 
  D.~T.~Son and M.~A.~Stephanov,
  Phys.\ Rev.\ Lett.\  {\bf 86}, 592 (2001)
  [hep-ph/0005225].

\end{thebibliography}
\end{document}